\documentstyle[aps,prl,epsf]{revtex}

\begin{document}

\title{Low-energy theorem for scalar and vector interactions of a composite
spin-1/2
system }

\author{S. J. Wallace$^{1}$, Franz Gross$^{2,3}$ and J. A.
Tjon$^{4}$}

\address{
$^{1}$ Department of Physics and Center for Theoretical Physics,
          University of Maryland, College Park, MD 20742 \\
$^{2}$Department of Physics,
College of William and Mary
Williamsburg, VA 23185  \\
$^{3}$CEBAF,
12000 Jefferson Avenue, Newport News, VA 23606, and  \\
$^{4}$ Institute for Theoretical Physics,
University of Utrecht, 3508 TA Utrecht, The Netherlands }
\date{\today}

\maketitle

\begin{abstract}
Scalar and vector interactions, with the scalar interaction coupled to a
composite spin-1/2 system so as to cause
a shift of its mass, are
shown to obey a low-energy theorem
which guarantees that the second order interaction due to z-graphs is the same
as for a point Dirac particle.  Off-shell and contact interactions appropriate
to
the composite system cancel and this is verified in a model of a composite
fermion.
The result provides a justification for the use of the Dirac
equation as it has been used in relativistic nuclear scattering and mean field
theories.
\end{abstract}

\twocolumn

The use of the Dirac equation in nuclear
physics has been a subject of interest
and debate in
recent years.  One of its outstanding successes
is in elastic scattering of
protons by nuclei. \cite{WallAnnRev,RayHoffCoker}
Quite large scalar and vector interactions,
which almost cancel one another, characterize
the proton-nucleus interaction.
Solving the Dirac equation with an attractive
scalar potential and a repulsive
vector one, each of magnitude about 300 MeV,
produces a good description of
spin observables at intermediate energies.
\cite{Clarketal1,MSW,SMW,Clarketal2,TjonWallIA2}
The principal effect is due to z-graphs when the Dirac equation
is used, but it
may be understood also at a simpler, classical level.

Consider a classical hamiltonian of the form
\begin{equation}
\label{masshift}
H = V + \sqrt{(M+S)^{2} + {\bf p} ^{2} },
\end{equation}
where $V$ is the time-component of a vector potential and $S$ is a scalar
potential, both
of which are taken to be spatially uniform.  Expanding in $S$ to second order
yields,
\begin{equation}
\label{pairterm}
H = \epsilon + V + \frac{M}{\epsilon} S + \frac{{\bf p} ^{2}}{2 \epsilon ^{3}}
S ^{2}
+ \cdots ,
\end{equation}
where $\epsilon = \sqrt{M ^{2} + {\bf p} ^{2}}$.
The momentum dependent repulsive potential term in
Eq.~(\ref{pairterm}) provides the main relativistic effect
in proton scattering by nuclei at intermediate energies.

To obtain the same results from the
Dirac equation with scalar and vector
potentials, one reduces the energy expression to the form,
\begin{equation}
\label{dirac}
E \psi = ( \epsilon + {\cal{V}} ^{++} + {\cal{V}} _{pair} )\psi ,
\end{equation}
where
\begin{equation}
{\cal{V}} ^{++} = V + \frac{M}{\epsilon} S
\end{equation}
 and
\begin{equation}
\label{Vpair}
{\cal{V}}_{pair} = \frac{{\bf p} ^{2} S ^{2} }
{\epsilon ^{2}[E + \epsilon - V + (M / \epsilon )S]}  =
\frac{{\bf p} ^{2} S ^{2}}{2 \epsilon ^{3}} + \cdots ,
\end{equation}
is the z-graph contribution.  To second order it is the same as in
Eq.~(\ref{pairterm}).
For a composite system such as a nucleon, general arguments have been
presented that z-graphs should be suppressed strongly.
\cite{Brodsky1,BrodskyBlesz,BrodskyJaros,Cohen}
Thus the correctness of ${\cal{V}}_{pair}$
as obtained from the Dirac equation for a nucleon is in doubt,
notwithstanding the classical basis for the effect.

In this Letter, we show that there is a
low-energy theorem for both scalar and vector interactions which
guarantees the correctness
of the
second order result Eq.~(\ref{Vpair}).
The result for the vector interaction is a straightforward generalization
of the low-energy theorem familiar from Compton scattering
\cite{Low,GellMannGoldberger}
(even though the vector interaction here is purely longitudinal), but the
result for a scalar interaction is a consequence of the observations that
(i) an external scalar interaction $S$ may be defined which acts to shift
the mass of any composite system from $M$ to $M + S$, and (ii) the scalar
interaction so defined satisfies Ward identities which quarantee that the
second order potential for the scattering of any composite spin-1/2 bound
state gives the universal result (\ref{Vpair}) at low energy.

To show explicitly how the low-energy theorem emerges for the scalar
interaction,  consider the following simple model lagrangian
\begin{equation}
\label{eq:lagrangian}
{\cal{L}} = \bar{\psi} ( i \partial \!\!\!/ - m )\psi
+ \frac{1}{2} [ \partial _{\nu} \phi \partial ^{\nu} \phi -
\mu ^{2} \phi ^{2}]  - g \bar{\psi} \psi \phi ^{2}.
\end{equation}
Assume that this system has a bound state of mass $M$ and spin-1/2
which is composed of the
elementary fermion of mass $m$ and the scalar boson of mass $\mu$.
By scaling all parameters in the lagrangian with dimensions of mass to
new values $m \rightarrow  \lambda m$, $\mu \rightarrow
\lambda \mu$, $g \rightarrow  \lambda ^{-1}g$, and
similarly scaling all cutoff or renormalization masses associated
with the theory, it is clear that the bound state mass $M$
will be scaled to $\lambda M$.  To obtain a scalar interaction which
satisfies the requirement (i) above, choose $\lambda =
1 + S/M$,
which implies that the lagrangian which includes interactions of the
scalar field $S$ has the form (to first order in $S$)
\begin{equation}
{\cal{L}} \rightarrow {\cal{L}}  - \rho S,
\end{equation}
where the scale-breaking charge associated with the mass scaling
is
\begin{equation}
\label{eq:rho}
\rho = (m/M)\bar{\psi} \psi
+ 2(\mu/M) \phi ^{2} -  (g/M) \bar{\psi} \psi \phi ^{2} .
\end{equation}
In general, this scale-breaking charge $\rho$ is proportional to the
divergence of the dilatation current of the system, including any anomalous
contributions generated by scaling of the cutoff masses required to
regularize the model.

The assumption that the scalar interaction is given by the scale-breaking
charge $\rho$ allows us to obtain a Ward identity for the vertex function
for the interaction of the scalar field with the composite fermion,
$\Lambda^S(p',p)$.  Before obtaining this, recall the Ward identity
for the vertex function for the vector interaction, $\Lambda^0(p',p)$, which
is
\begin{equation}
\label{Ward1}
\Lambda ^{0} (p,p) = - \frac{\partial \Sigma (p)}{\partial p _{0}} = -
A_{0} \gamma ^{0} - 2 p ^{0} (A'_{0} p\!\!\!/ + B'_{0}) ,
\end{equation}
where the self energy, $\Sigma (p)$, has the general form
\begin{equation}
\Sigma (p) = A(p^{2}) p\!\!\!/ + B(p^{2}) ,
\end{equation}
with $A$ and $B$ scalar functions, and $A _{0} = A(M^{2})$,  $A'_{0}
= dA(p^{2})/d p^{2}| _{p^{2}=M^{2}}$, and similarly for $B'_{0}$.  Although
a detailed expression
for $\Sigma (p)$ is not required, it is straightforward to obtain one
from the lagrangian of Eq.~(\ref{eq:lagrangian}), in which case $\Sigma$
is obtained from a loop graph involving a fermion and a scalar meson
propagator.  Note that only the time component of the current is needed.

The Ward identity for the scalar vertex has a form similar to
Eq.~(\ref{Ward1}).  Examination of the lowest order
Feynman diagrams in our simple model shows that the vertex function
corresponding to insertion of the charge $\rho$, in the
limit where $q=p'-p\to0$, is
\begin{eqnarray}
\label{scalarvertex1}
 \Lambda ^{S} (p,p) = \frac{m}{M} \frac{\partial \Sigma (p)}{\partial m
} +  \frac{\mu}{M} \frac{\partial \Sigma (p)}{\partial \mu } + \frac{g}{M}
\frac{\partial \Sigma(p)} {\partial g }  \nonumber \\ + \frac{\Lambda}{M}
\frac{\partial \Sigma (p)}{\partial \Lambda },
\end{eqnarray}
where $\Lambda$ is the cutoff mass.  Using the fact that
$\Sigma$ is dimensionless, and hence invariant when all
parameters with the dimensions of mass are scaled, e.g.,
\begin{equation}
\label{nondim}
\Sigma (m,\mu,\Lambda,g,p _{\alpha}) = \Sigma (\lambda m, \lambda \mu,
\lambda \Lambda, \lambda ^{-1} g, \lambda p _{\alpha})  ,
\end{equation}
and expanding to first order about $\lambda = 1$, one finds
\begin{equation}
\label{scalarvertex2}
\Lambda ^{S} (p,p) = -\frac{p _{\alpha}}{M} \frac{\partial \Sigma
(p)}{\partial p _{\alpha}} = - A_{0} \frac{p\!\!\!/}{M} - \frac{2 p
^{2}}{M} (A'_{0} p\!\!\!/ + B'_{0}) .
\end{equation}
This equation is a direct consequence of a
Ward identity for the divergence of the dilatation current \cite{Coleman}
and the low-energy theorem which we will derive depends on the
existence of such an identity.

We are now ready to use the identity (\ref{scalarvertex2}) to prove the
low-energy theorem for the scalar and vector interaction.  The propagator
of the composite spin-1/2 system  of mass $M$  and four-momentum $p$ may be
written
\begin{equation}
\label{Gofp}  G(p) =  \frac{1}{1 - \Sigma (p)}.
\end{equation}
By assumption, there is a pole in $G(p)$ at $p\!\!\!/ = M$.
Expanding the propagator about the bound-state pole, one finds
\begin{equation}
\label{prop}
G(p) =  Z _{2} \Bigl[ \frac{1}{ p\!\!\!/  - M } + \delta G(p) \Bigr],
\end{equation}
where
\begin{equation}
Z _{2} \equiv - \{ A_{0} + 2 M [ M  A'_{0} + B'_{0} ]
\}^{-1}=-(\Sigma'_0)^{-1} ,
\end{equation}
is a wave function normalization factor with $\Sigma'_0=d\Sigma(p\!\!\!/)/
dp\!\!\!/|_{p\!\!\!/=M}$.
Nonelementary propagation due to excited states of
invariant masses greater than $M$ gives rise to
\begin{eqnarray}
\delta G(p) =  Z_2\,{M+ p\!\!\!/\over 4M}\Sigma''_0 +
 D_{0} (  p\!\!\!/ - M)  + \cdots , \nonumber
\end{eqnarray}
where $\Sigma''_0=d^2\Sigma(p\!\!\!/)/
dp\!\!\!/^2|_{p\!\!\!/=M}$ and terms omitted from the expansion are higher
order in $p ^{2} - M ^{2}$ and do not play a role in the low-energy limit.
The detailed form of the
function $D_{0}$ is not required for the proof of the low-energy theorem
involving scalar interactions.
The usual spectral expansion of the
positive-energy pole term in $G(p)$ shows that the ground state of the
composite
system has the positive-energy wave function $ Z _{2} ^{1/2} u(p)$,
where $u(p)$ is a Dirac spinor for an elementary fermion of
mass $M$.

To simplify the notation, we define a vertex which combines the
scalar and vector interactions and coupling strengths as follows,
\begin{equation}
\Lambda ^{SV} (p,p) = S \Lambda ^{S} (p,p) + V
\Lambda ^{0} (p,p) .
\end{equation}
Forward scattering of the composite fermion from the scalar and
vector fields is studied in second order and
in the limit $q \rightarrow
0$, where $q$ is the momentum exchanged with the source.
The diagrams of Figure 1 yield for the potential ${\cal{V}}$ the
following expression,
\begin{eqnarray}
\label{pot2}
{\cal{V}}=
\frac{M}{2 \epsilon}Z _{2}^{1/2}&& \bar{u} (p) \Biggl\{
\Lambda ^{SV} (p, {p+q}) G({p+q}) \Lambda ^{SV} ({p+q},p)
\nonumber  \\ && + (q\to -q) + C ^{SV} (p,p) \Biggr\}
Z _{2} ^{1/2} u(p) \,  ,
\end{eqnarray}
where the first term in the curly braces is the direct pole term,
Fig.~1a, the second term with $q\to -q$ the crossed pole term, Fig.~1b, and
the third, contact-like term, Fig.~1c describes processes involving
scattering from the constituents within a single self-energy bubble. Note
that the iteration of the first order potential
is contained in Eq.~(\ref{pot2}).   It must be subtracted to avoid double
counting.  The appropriate subtraction is based on the second-order
scattering by the equivalent potential, $S + \gamma ^{0} V$, using the
positive-energy pole part of the Dirac propagator.

For time-like vector interactions at zero momentum transfer,
the contact term is $V ^{2}C ^{0 0} (p,p) = V ^{2}\partial ^{2}
\Sigma (p) / \partial p^{0} \partial p^{0}$.  For scalar
interactions the contact term corresponds to two insertions of
Eq.~(\ref{eq:rho}).
By expanding Eq.~(\ref{nondim}) to second order about $\lambda =
1$, it is possible to show that the interaction vertex to second order in
$ \rho$ is
\begin{eqnarray}
\label{SScontact}
C^{SS}(p,p) = &&(p _{\mu} p _{\nu} /
M^{2} ) \partial ^{2}\Sigma(p) / \partial p_{\mu} \partial
p_{\nu}\nonumber\\
&&\qquad\qquad + (2/M^{2}) p _{\mu}\partial\Sigma(p) / \partial p_{\mu}\, .
\end{eqnarray}
There are also cross terms involving one $\rho$ and one vector insertion.
Collecting the various contact terms and coupling strengths, we have,
\begin{eqnarray}
C ^{SV} (p,p) &=&  S^{2}C ^{SS} (p,p) + 2 S V C^{0 S }
(p,p) \nonumber +
V^{2} C ^{00} (p,p)
\end{eqnarray}
where
 $C ^{0 S} (p,p) = - \partial \Lambda ^{S} (p,p)/
\partial p _{0} $.

Because a denominator in $G(p \pm q)$ vanishes with $q$, it is necessary
to evaluate numerator factors correct to first order in
$q$ before going to the limit $q \rightarrow 0$.  This involves
expanding the vertex $\Lambda ^{SV} (p',p)$ about $p$ or $p'$ in a Taylor's
series,
\begin{eqnarray}
\label{vertexexpan}
\Lambda ^{SV} (p, p+q) = \Lambda ^{SV} (p,p) &&+ q _{\mu} \Bigl[
\frac{\partial \Lambda ^{SV} (p', p)}
{\partial p _{\mu}} \Bigr]
_{p' = p} \nonumber\\
&&+ \cdots  ,
\end{eqnarray}
and a similar expansion of $\Lambda ^{SV} (p-q, p)$.
Due to symmetries in the expansion, these contributions can be
expressed also as second derivatives of $\Sigma (p)$ with respect to
momenta.  Thus there are cancellations with the contact terms.

Finally, the $SS$ contribution to the potential (\ref{pot2})
follows from substituting (\ref{scalarvertex2}), (\ref{prop}),
(\ref{SScontact}), and (\ref{vertexexpan}).
Keeping only terms which contribute as $q \rightarrow 0$,  we find
\begin{equation}
\label{VSVcancel}
{\cal{V}} = \frac{{\bf{p}}^{2}S ^{2}}{2 \epsilon ^{3}} + {S^2\over \epsilon}
\bigr[\xi +(\xi-1)+(1-2\xi)
\bigr]  = \frac{{\bf{p}}^{2}S ^{2}}{2
\epsilon ^{3}} ,
\end{equation}
where $\xi=Z_2M\Sigma''_0/2$.
The contributions to Eq.~(\ref{VSVcancel}) arise as
follows: the ${\bf{p}}^{2}$ term from the composite particle z-graphs,
$\xi$ from the off-shell propagation $\delta G(p)$, $\xi-1$ from the contact
terms, and $(1-2\xi)$ from the off-shell expansion of the vertex functions,
Eq.~(\ref{vertexexpan}).
Cancellations render the overall result
independent of the factor $\xi$.
This demonstrates the low-energy theorem
for the scalar interaction given in
Eq.~(\ref{eq:rho}), and shows that a
scalar interaction capable of shifting the mass
generates a repulsive
potential of the same form as that obtained classically
from a mass shift in Eq.~(\ref{masshift}) or from the
Dirac equation from the z-graph contribution,
 ${\cal{V}} _{pair}$.

We have carried out a similar analysis for the $VV$ and $SV$ terms and
find that the second order potential is zero, in agreement with
Eq.~(\ref{Vpair}).  This result emerges from the cancellation of five
terms: the four which arose in the scalar case plus a new term.  This
new term appears because
of the subtraction mentioned below Eq.~(\ref{pot2}).

The low-energy theorem establishes an equivalence between second
order scattering of a composite spin-1/2 system by scalar
plus vector interactions of arbitrary strength and the second-order scattering
of a Dirac
particle by similar potentials.  The key condition is that the scalar
interaction be coupled to the scale-breaking charge so as to cause a
mass shift.
The composite particle
z-graph contributions due to such a scalar interaction
are not suppressed.

The analysis suggests conditions under which the
relativistic effect in
proton-nucleus scattering may be unaffected by compositeness of the
nucleon.
Whether these conditions
apply to QCD remains an open question.

\vspace{0.2in}

This work is was initiated at a joint CEBAF/INT workshop,
September 17-26, 1993.
Support by the U.S. Department of
Energy under grants DE-FG02-93ER-40762 (S.J.W.) and
DE-AC05-84ER40150 (F.L.G.) is gratefully acknowledged.


\begin{references}
%
\bibitem{WallAnnRev}S. J. Wallace, Ann. Rev. Part. and Nucl. Sci.
{\bf{36}}, 267 (1987).
%
\bibitem{RayHoffCoker}L. Ray, G. W. Hoffmann and W. R. Coker,
Physics Reports {\bf{212}}, 223 (1992).
%
\bibitem{Clarketal1},B. C. Clark, L. Mercer, and P. Schwandt,
 Phys. \  Lett. {\bf{122B}}, 211 (1983).
%
\bibitem{MSW}J. A. McNeil, J. R. Shepard and S. J. Wallace,
Phys. \ Rev. \ Lett. {\bf{50}}, 1439 (1983).
%
\bibitem{SMW}J. R. Shepard, J. A. McNeil and S. J. Wallace,
Phys. \ Rev. \ Lett. {\bf{50}}, 1443 (1983).
%
\bibitem{Clarketal2}B. C. Clark, S. Hama, R. L. Mercer, L. Ray and B. D. Serot,
Phys. \ Rev. \ Lett. {\bf{50}}, 1643 (1983).
%
\bibitem{TjonWallIA2}J. A. Tjon and S. J. Wallace,
Phys. \ Rev. \ {\bf{C36}}, 1085 (1987).
%
\bibitem{Brodsky1}S. J. Brodsky, Comments Nucl. and Part. Phys. {\bf{12}},
213 (1984).
%
\bibitem{BrodskyBlesz}E. Bleszynski, M. Bleszynski and S. Brodsky,
Phys. \ Rev. \ Lett. {\bf{59}}, 423 (1987).
%
\bibitem{BrodskyJaros}T. Jaroszewicz and S. J. Brodsky,
Phys. \ Rev. \ {\bf{C43}}, 1946 (1991).
%
\bibitem{Cohen}T. D. Cohen, Phys. \ Rev. \ {\bf{C45}}, 833 (1992).
%
\bibitem{Low}F. E. Low, Phys. \ Rev. \ {\bf{96}}, 1428 (1954).
%
\bibitem{GellMannGoldberger}M. Gell-Mann and M. L. Goldberger,
Phys. \ Rev. \ {\bf{96}}, 1431 (1954).
%
\bibitem{Coleman}S. Coleman, {\it{Aspects of Symmetry}}, (Cambridge Univ.
Press, New
York, 1988), p. 67.
%
\end{references}
\end{document}